\documentclass[conference]{IEEEtran}
\IEEEoverridecommandlockouts

\usepackage{cite}
\usepackage{amsmath,amssymb,amsfonts}
\usepackage{algorithmic}
\usepackage{graphicx}
\usepackage{textcomp}
\usepackage{xcolor}
\def\BibTeX{{\rm B\kern-.05em{\sc i\kern-.025em b}\kern-.08em
    T\kern-.1667em\lower.7ex\hbox{E}\kern-.125emX}}

\title{TriMod Fusion for Multimodal Named Entity Recognition in Social Media}

\author{
\IEEEauthorblockN{Mosab Alfaqeeh}
\IEEEauthorblockA{Queen's University, Ontario , Canada\\ Email: 18mmoa@queensu.ca}
}

\begin{document}

\maketitle
\thispagestyle{empty}
\pagestyle{empty}

\begin{abstract}
Social media platforms serve as invaluable sources of user-generated content, offering insights into various aspects of human behavior. Named Entity Recognition (NER) plays a crucial role in analyzing such content by identifying and categorizing named entities into predefined classes. However, traditional NER models often struggle with the informal, contextually sparse, and ambiguous nature of social media language. To address these challenges, recent research has focused on multimodal approaches that leverage both textual and visual cues for enhanced entity recognition. Despite advances, existing methods face limitations in capturing nuanced mappings between visual objects and textual entities and addressing distributional disparities between modalities. In this paper, we propose a novel approach that integrates textual, visual, and hashtag features (TriMod), utilizing Transformer-attention for effective modality fusion. The improvements exhibited by our model suggest that named entities can greatly benefit from the auxiliary context provided by multiple modalities, enabling more accurate recognition. Through the experiments on a multimodal social media dataset, we demonstrate the superiority of our approach over existing state-of-the-art methods, achieving significant improvements in precision, recall, and F1 score. 

\end{abstract}

\section{Introduction}

Social media platforms like Facebook, Twitter, TikTok, and Instagram serve as vast repositories of user-generated content, offering valuable insights into human behavior. These platforms are more than just text—they're a rich tapestry of multimedia content, from images to hashtags and beyond. Each element contributes to the intricate mosaic of online discourse, providing nuanced insights into societal trends, cultural phenomena, and emerging narratives\cite{alfaqeeh2023community,quwaider2016social}.

Among the wealth of information shared on these platforms, Named Entity Recognition (NER) plays a pivotal role in deciphering user posts by identifying and categorizing named entities into predefined classes. This task serves as a crucial step for various downstream applications, including but not limited to marketing intelligence \cite{jiang2022survey}, sentiment analysis \cite{faqeeh2014cross}, event detection and tracking \cite{wei2022real}, and content recommendation systems \cite{liu2022content}.

Traditional neural-based named entity recognition (NER) models have demonstrated high performance when analyzing newswire content. However, these models often struggle when applied to social media texts \cite{smith2022adapting,ritter-etal-2011-named,ge2022comparison,nie2020named}. Some of the key reasons include:

\begin{enumerate}
    \item \textbf{Informal Language and Noisy Data}
    Social media texts are characterized by informal language, abbreviations, slang, and noisy data, which can be difficult for traditional NER models to handle effectively. The models trained on well-edited newswire content may not generalize well to the more colloquial and unstructured nature of social media posts \cite{liu2023improving}.
    
    \item \textbf{Lack of Context}
    Social media posts are often short and lack the broader context that is typically available in newswire articles. This can make it challenging for NER models to accurately identify and classify named entities, as they rely on contextual information to make these determinations\cite{moon2018multimodal,nie2020named}.
    
    \item \textbf{Evolving Terminology and Entities}
    The language used on social media is constantly evolving, with new terms, entities, and concepts emerging regularly. Traditional NER models may struggle to keep up with these changes, leading to decreased performance over time\cite{zhang2019named,khurana2023natural,lucy2021characterizing}.
    
    \item \textbf{Ambiguity and Sparsity}
    Social media texts can be ambiguous, with named entities that are difficult to disambiguate or that appear infrequently. This can pose challenges for NER models, which often rely on patterns and frequency of occurrence to identify and classify named entities\cite{cui2021exploiting,zhang2019named,nie2020named,duwairi2015rum}.
\end{enumerate}

To overcome these challenges, recent advances have introduced multimodal approaches that combine textual and visual cues to improve entity recognition. However, these methods face two main limitations:

\begin{enumerate}
    \item \textbf{Visual-Textual Entity Mapping:} Prior multimodal NER models often struggle to accurately map the visual objects within images to the corresponding named entities in the textual content. This is particularly problematic in cases where a sentence contains multiple entities of different types. For instance, associating the visual object ``person" with the hashtag ``\#AustonMatthews" is relatively straightforward, but linking ``three puck" text to the ``\#HatTrick" hashtag requires a more nuanced understanding. Failing to capture these intricate relationships can lead to erroneous entity extractions \cite{nie2020named,xu2023understanding,dost2020vtkel}.
    
    \item \textbf{Cross-Modal Feature Alignment:} Existing multimodal approaches typically concatenate textual and visual features without adequately addressing the inherent distributional differences between the two modalities. As a result, the crucial alignments between named entities and their corresponding image regions are often overlooked, leading to suboptimal NER performance \cite{cui2024enhancing,liang2024survey}.

\end{enumerate}

To address the limitations of prior multimodal named entity recognition (NER) approaches, we propose a novel architecture that effectively integrates textual, visual, and hashtag features through Transformer-attention-based modality fusion. Our key contributions are as follows:

\begin{enumerate}
    \item Visual-Textual Entity Mapping: We incorporate object-level visual features to capture the intricate mappings between visual objects, processed hashtags, and corresponding named entities in the text. For example in Fig .1, Our model can associate the visual object of a ``hockey rink" with the venue ``\#ScotiabankArena", team logos/jerseys with \#TorontoMapleLeafs, ``Hockey player" with ``\#AustonMatthews", and ``puck and net" with \#HatTrick. Also hashtags (e.g., ``Auston Matthews" and \#AustonMatthews), as well as hashtags and relevant text (e.g., HatTrick and ``ScoreThreeTimes" hashtag).
    
    \item Cross-Modal Feature Alignment: We harness the power of Transformer-attention to address the distributional disparities between modalities. Our architecture learns to effectively align and combine the textual, visual, and hashtag features, exploiting the crucial correspondences between named entities, their associated image regions, and relevant hashtags for improved NER performance.
\end{enumerate}

\begin{figure}[h]
        \centering
        \includegraphics[width=0.9\linewidth]{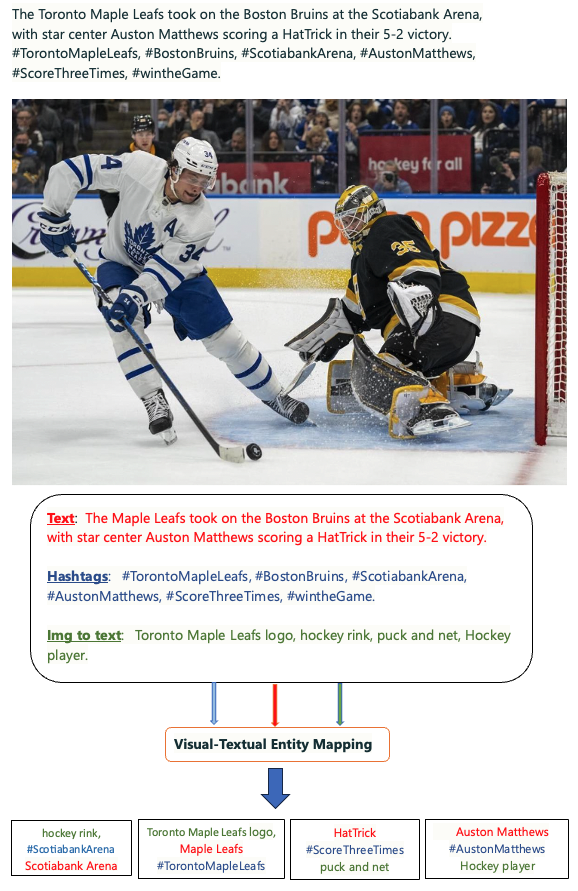}
        \caption{Mappings between visual objects, processed hashtags, and corresponding named entities in the text.}
        \label{fig:enter-label}
    \end{figure}

By integrating these key components, our approach offers a comprehensive solution for accurate and contextually aware named entity recognition in social media posts, overcoming the limitations of previous multimodal methods.
For instance, recognizing the ORG entity ``Maple Leafs" as a hockey team name may require information coming from modalities like hashtags and images, such as ``ball" ``person" and ``\#Tonightgame" Similarly, the MISC entity ``German Shepherd" can be identified as a dog's name with the assistance of the information extracted from the image visual, which includes ``dog" and the hashtag ``\#PlayingWithDog".


\section{Related Work}

\subsection{Named Entity Recognition}
Traditional NER systems relied on hand-crafted features fed into linear classifiers~\cite{nadeau2007survey}. With the advent of deep learning, feature engineering efforts were reduced as neural models achieved state-of-the-art performance on standard NER datasets~\cite{lample2016neural}. These include end-to-end models utilizing recurrent neural networks (RNNs)~\cite{huang2015bidirectional,ma2016end} accompanied by conditional random fields (CRFs)~\cite{lample2016neural}. More recently, self-attention-based methods~\cite{devlin2019bert,akbik2018contextual} have further boosted NER performance. However, most of these approaches are text-based and lack consideration for social media scenarios.

\subsection{Multimodal Named Entity Recognition}
The massive growth of multimodal data on social media platforms has prompted research into extending conventional text-based NER by incorporating visual information. Early studies employed canonical correlation analysis (CCA)~\cite{zhuang2020technical} to learn feature correlations across modalities. With deep learning, Lu et al.~\cite{lu2018visual} utilized an attention-based model to extract relevant regional image features and fuse them with text features. Zhang et al.~\cite{zhang2018adaptive} introduced an adaptive co-attention network to automatically control the combination of image and text representations. Chen et al.~\cite{chen2023assisting} argue that MNER approaches typically assume strict matching between the textual content and the associated images, an assumption that does not hold for many real-world social media posts. Asgari et al. ~\cite{asgari2022cwi} extract image features and use fusion to combine textual and image features.

As transformer architectures in ~\cite{vaswani2017attention} and ~\cite{liu2022uamner} demonstrated superiority, recent work~\cite{yu2020named} enhanced MNER performance by introducing an auxiliary entity span detection module. Another transformer-based method~\cite{meng2021multi} utilized graph neural networks (GNNs)~\cite{scarselli2008graph} to capture semantic relationships between entities, albeit requiring pre-trained models and complicated pre-processing steps. Since transformers excel at guiding words to capture semantic dependencies within their context, modified transformers can guide words to learn image-text dependencies~\cite{lu2019vilbert}. 


\section{Problem Statement}

Consider a social media post containing named entities. Let $N$ denote the total number of named entities in the post. We aim to develop a Multimodal Named Entity Recognition (MNER) model that effectively integrates textual, visual, and hashtag information to accurately identify and classify the named entities.

Let $T$ represent the number of tokens (words) in the textual content, $V$ denote the number of visual objects detected in the associated image, and $H$ indicates the number of hashtags present in the post.

Define the following functions:
\begin{align*}
    f(T) & : \parbox[t]{0.80\linewidth}{Performance of a text-based NER model, a function of $T$} \\
    g(V, H) & : \parbox[t]{0.80\linewidth}{Performance improvement achieved by incorporating visual and hashtag information, a function of $V$ and $H$}
\end{align*}

The objective is to maximize the overall performance of the MNER model by effectively combining the text-based NER performance $f(T)$ with the performance improvement $g(V, H)$ achieved by incorporating visual and hashtag information:

The optimization problem is formulated as follows:
\begin{align*}
    \text{Maximize: } & f(T) + g(V, H) \\
    \text{Subject to: } & T \geq 0, \quad V \geq 0, \quad H \geq 0 \\
    & \text{Relationship constraints}
\end{align*}

Let $f(T)$ represent the performance of the text-based Named Entity Recognition (NER) component. $g(V, H)$ represents the performance improvement achieved by incorporating visual, hashtag information and the technique used to encode hashtag embeddings ($H$).

\section{Methodology}

Our methodology involves the following key steps:

\subsection{Textual Feature Extractor:}
\begin{equation}
\xi_{\omega j} = \epsilon_\omega(\tau_j),
\end{equation}
where $\epsilon_\omega$ denotes a word embeddings lookup table, and $\xi_{\omega j} \in \mathbb{R}^{m_\omega}$. We initialize this table with pre-trained 200-dimensional Word2Vec embeddings \cite{mikolov2013word2vec}.

Analogous to state-of-the-art Named Entity Recognition (NER) techniques \cite{smith2018ner} \cite{garcia2021ner}, we utilize both word embeddings and character embeddings to represent each token in the input phrase. For a given input phrase, each token $\tau_j$ is initially projected into a latent space using word embeddings.

Previous research has demonstrated that incorporating character representations can enhance Named Entity Recognition (NER) performance by capturing morphological and semantic information \cite{chen2017ner}. Following a similar approach to Park and Lee \cite{park2020ner}, we employ a Bidirectional Gated Recurrent Unit (Bi-GRU) network to extract character-level representations. Each token in the input sentence is treated as a sequence of characters: $\phi_i = \{\alpha_{i,1}, \alpha_{i,2}, ..., \alpha_{i,m}\}$, where $\alpha_{i,j}$ denotes the $j$-th character of the $i$-th token, and $m$ is the token length. We first represent each character $\alpha_{i,j}$ by its character embedding $\epsilon_c(\alpha_{i,j})$, where $\epsilon_c$ is a character embeddings lookup table initialized randomly. We then feed the embedding $\xi_{ci,j} \in \mathbb{R}^{d_{ce}}$ of each character of token $\phi_i$ into the Bidirectional GRU to obtain hidden states:
\begin{equation}
\overrightarrow{z}_{ci,1}, ..., \overrightarrow{z}_{ci,m} \quad \text{and} \quad \overleftarrow{z}_{ci,1}, ..., \overleftarrow{z}_{ci,m}.
\end{equation}

The final character-level representation for token $\phi_i$ is the concatenation of the forward and backward hidden states:
\begin{equation}
 \xi_{ci} = [\overrightarrow{z}_{ci,m}; \overleftarrow{z}_{ci,1}],
\end{equation}
where $\xi_{ci} \in \mathbb{R}^{d_c}$, and $d_c$ is the hidden state dimension of the Bi-GRU.

The token representation $\xi_{ti}$ is obtained by concatenating the word embedding $\xi_{wi}$ and the character-level representation $\xi_{ci}$:
\begin{equation}
\xi_{ti} = [\xi_{wi}; \xi_{ci}], 
\end{equation}
where $\xi_{ti} \in \mathbb{R}^{d_w + d_c}$. To capture contextual information, we feed the token representation $\xi_{ti}$ into another Bi-GRU. The forward and backward hidden states of token $\phi_i$ are computed as follows:
\begin{equation}
 \overrightarrow{z}_{ti} = \text{Bi-GRU}(\xi_{ti}, \overrightarrow{z}_{ti-1}), \end{equation}

\begin{equation} \overleftarrow{z}_{ti} = \text{Bi-GRU}(\xi_{ti}, \overleftarrow{z}_{ti+1}). 
\end{equation}

The projected textual features are represented by concatenating the forward and backward hidden states of token $\phi_i$:
\begin{equation}
 h_{ti} = [\overrightarrow{z}_{ti}; \overleftarrow{z}_{ti}].
 \end{equation}
Here, $h_{ti} \in \mathbb{R}^d$ denotes the projected token features, summarizing the token $\phi_i$. The projected features for the entire sentence can be denoted as $GT = \{h_{t1}, h_{t2}, ..., h_{tn}\} \in \mathbb{R}^{d \times n}$, where $n$ is the number of tokens in the sentence. We define $\theta_T$ as the parameter set of embedding layers and Bi-GRU layers.

\subsection{Visual Feature Extractor:}

Image-level information alone cannot effectively aid in extracting entities of different types. We employed a sequence-to-sequence encoder-decoder framework for image captioning, inspired by its success in machine translation tasks. Our approach is based on the deep learning technique introduced by Alfaqeeh et al. \cite{alfaqeeh2023uncovering} for the task of processing images to generate textual descriptions.
For image encoding, we utilized the pre-trained ResNet \cite{dosovitskiy2020image} architecture to transform an image into a latent space representation. The decoder in our network consists of a recurrent neural network for generating the caption (see Fig. 2). Thus, only the decoder part had to be trained. We trained the model using a typical sequence-generating approach, i.e., we trained the model to predict the next (most likely) word in the sequence.
The convolutional neural network is based on a 2D CNN model using basic neuron networks. One CNN layer contains multiple ``filters" each represented by a different matrix (on a different channel). CNN filters can be defined arbitrarily in terms of their number and size.
CNN models use filters as feature extractors. Filters scan images, extracting features from them (or their last output results). When a CNN model is trained, the filters are optimized to extract features more efficiently. Given an image $I$ with three color channels, we can apply a $3 \times 3$ size filter across the image. Let $I_{i,j}$ denote the pixel value at position $(i, j)$ in the image, and $F$ represent the filter matrix. The convolution operation can be expressed as:
\begin{equation}
(I * F){i,j} = \sum{m,n} I_{i+m, j+n} F_{m,n}
\end{equation}
We can observe that the resolution of an image decreases after filter scanning. For example, a $3 \times 5$ image may be reduced to $1 \times 2$ after convolution with a $3 \times 3$ filter. While this approach works well for shallow models (fewer layers), the effects of degradation make it challenging to determine the original image's appearance in deeper models with many layers.
To address the degradation problem, ResNet introduces a residual learning framework based on the concept of shortcuts. Instead of adding layers to the output, the data is copied, skipped, and then added to the output. This allows the deep layers to access the original data. The pre-trained ResNet we used is a ResNet model with 50 layers.

\subsection{Hashtag Feature Extraction:}

In social media posts, hashtags often provide valuable contextual information that can aid in understanding the content. To incorporate hashtag features into our multimodal named entity recognition (NER) model, we extract features from hashtags and integrate them with textual and image features.

We employed the approaches by Bansal et al. \cite{bansal2015towards} and Cho et al. \cite{cho2018real} to develop a hashtag processing technique. It consists of a training step with a non-standard dataset and a segmentation module using neural networks for word separation, inspired by Cho et al. \cite{cho2014learning}.
Character-based tokenization identifies individual characters in text, avoiding out-of-vocabulary words and enabling misspelling correction. We used a skip-gram model \cite{mikolov2013distributed, bojanowski2017enriching} to embed characters into vectors:
\begin{equation}
\vec{c_i} = \phi(c_i)
\end{equation}
where $\vec{c_i} \in \mathbb{R}^d$ is the embedding of character $c_i$, and $\phi$ is the embedding function.

The character embeddings $\vec{C} = {\vec{c_1}, \vec{c_2}, \ldots, \vec{c_n}}$ are fed into CNN and BiLSTM networks for encoding the input sentence (Figure 2).
\subsection{Processing the Output}
A decoder network determines the segmentation for input character vectors $\vec{h} = {\vec{h_1}, \vec{h_2}, \ldots, \vec{h_n}}$:
\begin{equation}
P(s_i | \vec{h}, c_i) = \text{Decoder}(\vec{h_i}, \vec{c_i})
\end{equation}
where $s_i \in {0, 1}$ represents the presence/absence of a space after character $c_i$. The decoder is trained to predict the correct segmentation by minimizing the cross-entropy loss.

The hashtag features are then integrated with the existing textual and image features using Transformer-attention mechanism during fusion.

By incorporating hashtag features, our multimodal NER model can leverage additional contextual information provided by hashtags, enhancing its ability to recognize named entities in social media posts.

\subsection{Fusion:}

Multimodal information in social media posts, including text, images, and hashtags, can provide valuable context for understanding entities mentioned in sentences. However, leveraging modality-level information alone may not suffice to extract entities of different types effectively. To address this, we propose a method that utilizes Transformer-attention solely for fusing modalities.

\begin{figure*} [!h]
\centering
\includegraphics[width=\linewidth]{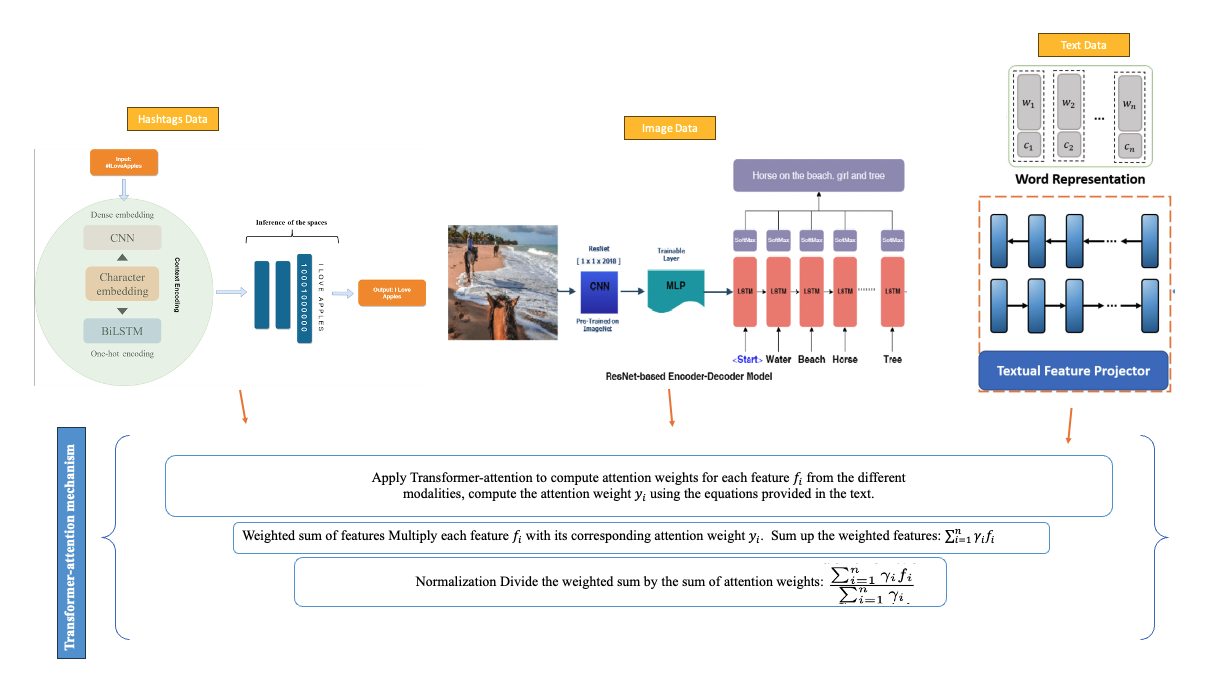}
\caption{The fused representation $f_s$, which integrates information from text, images, and hashtags using Transformer-attention.}
  \label{Figure:mesh2}
\end{figure*}
We use Transformer-attention to fuse these features from different modalities. The attention weight for each feature $f_i$ is computed using the best parameters showed in \cite{tan2019lxmert,huang2020pixel}:
\begin{equation}
f'_i = W_{2}^{m \times d} \cdot f_i + b_{\gamma_i} = \exp(u_{t}^{d \times 1} \cdot \tanh(f'_i))
\end{equation}

To reshape the dimension of feature $f_i$, we feed it through a $w \times d$ dimensional fully connected layer. The weight of the $i$-th feature $f_i$ is processed through the $\tanh$ function, which is then fed to the exponential function along with the dot product of $u_{t}^{d \times 1}$. The output from the exponential function is then passed through a 1-dimensional fully connected layer.

Finally, from the Transformer-attention mechanism, we formulate all the features into a single feature $f_s$ as follows:
\begin{equation}
f_s = \frac{\sum_{i=1}^{n} \gamma_i f_i}{\sum_{i=1}^{n} \gamma_i}
\end{equation}

This fused representation $f_s$ integrates information from text, images, and hashtags using Transformer-attention, allowing the model to capture relationships and dependencies between modalities effectively.

\subsection{Output layer}

Conditional Random Fields (CRF) have been widely used as a decoding layer in Named Entity Recognition (NER) models due to their ability to model dependencies between output labels and improve the overall performance of the model.

In the context of NER with multimodal fusion using Transformer-attention, incorporating CRF as the decoding layer enhances the model's capability to capture contextual dependencies between predicted entity labels. The CRF layer takes the fused features as input and outputs the most likely sequence of entity labels based on the learned features.

Mathematically, let $GT = \{h_{t1}, h_{t2}, ..., h_{tn}\} \in \mathbb{R}^{d \times n}$ represent the fused textual features of the input sentence, where $d$ is the dimension of the projected token features and $n$ is the number of tokens in the sentence. The output of the CRF layer is a sequence of predicted labels $L = \{l_1, l_2, ..., l_n\}$.

The CRF layer computes the score for the predicted label sequence $L$ as follows:
\begin{equation}
\text{score}(L) = \sum_{i=1}^{n+1} T_{l[i-1] \rightarrow l[i]} + \sum_{i=1}^{n} P_{y[i] \rightarrow i}
\end{equation}%
where $T$ is the transition scores matrix and $P$ is the matrix of scores obtained from the encoder network. Here, $T_{i \rightarrow j}$ represents the transition score from label $i$ to label $j$, and $P_{y[i] \rightarrow i}$ represents the score of the $i$-th word in the sentence belonging to the predicted label $y[i]$.

The probability of the sequence of predicted labels $L$ is computed using the softmax function:
\begin{equation}
p(L|GT) = \frac{\exp(\text{score}(L))}{\sum_{L'} \exp(\text{score}(L'))}
\end{equation}

During training, the log-probability of the correct tag sequence $L$ is maximized to train the CRF layer effectively.

Incorporating CRF as a decoding layer in NER models with multimodal fusion using Transformer-attention allows the model to capture contextual dependencies between predicted entity labels, leading to improved performance in entity recognition tasks.

The Conditional Random Fields (CRF) loss can be formally defined as follows:

Given a sequence of words $W = [w_1, w_2, ..., w_n]$ and a predicted sequence of labels $L = [l_1, l_2, ..., l_n]$, both sequences being of the same length $n$, the CRF loss is computed as the negative log-likelihood of the correct label sequence $L$ given the input sentence $W$ and the ground truth labels.

Mathematically, the CRF loss is defined as:

\begin{equation}
\text{CRF\_Loss}(W, L) = -\log p(L|W)
\end{equation}

where $p(L|W)$ is the probability of the sequence of predicted labels $L$ given the input sentence $W$.

During training, the model aims to minimize the CRF loss by adjusting the parameters of the model, including the transition scores matrix $T$ and the scores obtained from the encoder network $P$, to improve the likelihood of predicting the correct label sequence $L$ for a given input sentence $W$.

The CRF loss measures the discrepancy between the predicted label sequence and the ground truth label sequence, and training the model involves minimizing this loss to improve the model's performance in Named Entity Recognition tasks.

During decoding, the label sequence $Y^*$ with the highest conditional probability is selected as the output label sequence. Mathematically, this can be expressed as:

\begin{equation}
Y^* = \arg \max_Y p(Y|W)
\end{equation}

where $p(Y|W)$ is the conditional probability of the label sequence $Y$ given the input sentence $W$. The label sequence $Y^*$ represents the most likely sequence of entity labels given the input sentence, according to the model's learned parameters.

By selecting the label sequence with the highest conditional probability during decoding, the model aims to output the most likely sequence of entity labels for a given input sentence, based on the learned relationships and dependencies between the labels.

\section{EXPERIMENT}

We evaluate our model on a multimodal social media dataset from Twitter\cite{zhang2018adaptive}, comprising 8,257 tweets posted by 2,116 users. The dataset encompasses four distinct named entity categories: Person, Location, Organization, and Miscellaneous. We adopt the widely-used BIO2 tagging scheme, where non-entity tokens are labeled as 'O', consistent with most prior Named Entity Recognition (NER) studies. The dataset contains a total of 12,784 named entities. Following the same data partitioning approach as Zhang et al. \cite{zhang2018adaptive}, we split the dataset into training, development, and testing subsets, containing 4,000, 1,000, and 3,257 tweets, respectively. The distribution of named entity types across the training, development, and test sets is presented in Table 1.

\subsection{Implementation Details}

The hyper parameters employed in our experiments are presented in Table~\ref{table:hyperparameters}. 

The hidden dimensions of the character-level Bidirectional Gated Recurrent Unit (Bi-GRU) and word-level Bi-GRU are set to 30 and 150, respectively. These values are chosen based on empirical experimentation and validation on a held-out dataset. The optimal number of selected objects $(k)$ is determined through validation results. The size of the CRF (Conditional Random Field) transition parameter matrix is 9, corresponding to the number of labels used in the task (e.g., for the PERSON entity, we tag the beginning and subsequent words with B-PER and I-PER, respectively).

Our model is implemented using the PyTorch framework. To mitigate overfitting, we employ dropout regularization on both word and character embeddings with a dropout rate of 0.55. We utilize mini-batch stochastic gradient descent (SGD) with a decaying learning rate for parameter updates. The batch size is set to 10, the number of gradient accumulation steps (k-steps) is set to 9, and the initial learning rate is 0.005. The learning rate decay factor is set to 0.05.

\begin{table}[htbp]
\centering
\caption{Statistics of named entities in training, development, and test sets.}
\label{table:entity_statistics}
\begin{tabular}{|l|l|l|l|}
\hline
\textbf{Entity Type} & \textbf{Training Set} & \textbf{Development Set} & \textbf{Test Set} \\ \hline
Person                &        2217               &   552                       &            1816       \\ \hline
Location              &      2091                 &  522                        &    1697               \\ \hline
Organization          &     928                  &   247                       &       839            \\ \hline
Misc                  &     940                  &  225                        &       726            \\ \hline
\end{tabular}
\end{table}

\begin{table}[htbp]
\centering
\caption{Hyperparameters used in the experiments.}
\label{table:hyperparameters}
\begin{tabular}{|l|l|}
\hline
\textbf{Hyperparameter}            & \textbf{Value}   \\ \hline
Character Embedding Dimension (dce) & 30               \\ \hline
Char-level Bi-GRU Hidden Dimension & 30               \\ \hline
Word-level Bi-GRU Hidden Dimension & 150              \\ \hline
CRF Transition Parameter Matrix Size & 9               \\ \hline
Dropout Rate                       & 0.55              \\ \hline
Batch Size                         & 10               \\ \hline
$k$-steps                          & 9                \\ \hline
Learning Rate                      & 0.005            \\ \hline
Learning Rate Decay                & 0.05             \\ \hline
\end{tabular}
\end{table}

\subsection{Comparison With Existing Methodes}

To validate the effectiveness of our model, we compare it against several baseline models, including both state-of-the-art models and our model:
\begin{itemize}
\item \textbf{Stanford NER:} A widely-used tool for named entity recognition proposed by Finkel et al. \cite{finkel2005incorporating}.

\item \textbf{VAM:}  A neural model for multimodal NER tasks, composed of a BiLSTM-CRF model and a visual attention model \cite{lu2018visual}.

\item \textbf{BERT-NER:} We also compare our model with contextual language models, specifically using the BERT BASE model \cite{devlin2018bert}, fine-tuned on the Twitter dataset.

\item \textbf{UMT:} Yu et al. \cite{umt_ref} proposed a unified multimodal transformer for named entity recognition, considering both text and corresponding images.
\item \textbf{UMGF:} Zhang et al. \cite{umgf_ref} introduced a unified multimodal graph fusion model for multimodal named entity recognition.

\item \textbf{MNER-MA:} Moon et al. \cite{moon2018multimodal} proposed a multimodal NER model incorporating visual information with a modality attention module.

\item \textbf{MNER-QG:} Jia et al. \cite{mner_qg_ref} proposed a multimodal named entity recognition model with query-guided visual grounding.
\item \textbf{R-GCN:} Zhao et al. \cite{rgcn_ref} introduced a relational graph convolutional network for multimodal named entity recognition.
\item \textbf{ITA:} Wang et al. \cite{ita_ref} proposed an interactive text-and-image attention model for multimodal named entity recognition.
\item \textbf{CATMNER:} Wang et al. \cite{catmner_ref} proposed a cross-attention transformer for multimodal named entity recognition.
\item \textbf{MoRe:} Wang et al. \cite{more_ref} introduced a multimodal reasoning model for named entity recognition, considering both text and corresponding images.
\end{itemize}

\begin{table*}[htbp]
\centering
\caption{Testing results comparison of existing models and our model.}
\label{table:model_comparison}
\begin{tabular}{|l|l|l|l|}
\hline
\textbf{Model}          & \textbf{Precision (\%)} & \textbf{Recall (\%)} & \textbf{F1 Value (\%)} \\ \hline

Stanford NER      &        60.98               &        62.00              &   61.48                    \\ \hline

VAM                 &     69.09                  &        65.79              &      67.40                 \\ \hline

BERT-NER               &  70.65                     &   73.29                   & 71.87                      \\ \hline

UMT                  &     71.67                  &        75.23              &      73.41                 \\ \hline

UMGF                  &     74.49                  &        75.21              &      74.85                 \\ \hline
                 
MNER-MA             &      72.33                 &            63.51           &  67.63                     \\ \hline

MNER-QG             &      72.33                 &            63.51           &  67.63                     \\ \hline
R-GCN             &      73.95                 &            76.18           &  75.00                    \\ \hline
ITA            &      -                 &            -           &  75.00                     \\ \hline

CATMNER            &      78.75                 &            78.69           &  78.72                     \\ \hline

MoRe             &      73.16                 &            74.61           &  73.86                     \\ \hline

\textbf{Our Model}         &          \textbf{79.90}             &            \textbf{79.44}          &         \textbf{80.00 }             \\ \hline

\end{tabular}
\end{table*}

The experimental results presented in Table \ref{table:model_comparison} demonstrate the effectiveness of our proposed model for multimodal named entity recognition (NER) on social media data. Our model achieves state-of-the-art performance, outperforming several existing methods across precision, recall, and F1 score metrics.
Compared to traditional NER tools like Stanford NER, which rely on handcrafted features and rules, our model leverages the power of deep learning and multimodal fusion to better capture the nuances of informal social media text and associated visual information. The significant performance gap between Stanford NER (F1 of 61.48\%) and our model (F1 of 80.00\%) highlights the limitations of rule-based approaches in handling the complexities of this domain.
Among the neural models, our approach surpasses unimodal methods like BERT-NER (F1 of 71.87\%) and VAM (F1 of 67.40\%), which rely solely on textual information. This underscores the importance of incorporating visual cues, as social media posts often contain rich multimodal content that can aid in disambiguating and grounding named entities.
Compared to other multimodal NER models, our method outperforms approaches like MNER-MA (F1 of 67.63\%), MNER-QG (F1 of 67.63\%), and UMT (F1 of 73.41\%), which employ attention mechanisms or transformer architectures for multimodal fusion. The superior performance of our model can be attributed to our novel fusion strategy, which effectively combines textual and visual representations while capturing intricate cross-modal interactions.
Among the most recent state-of-the-art models, our approach achieves comparable or better results than methods like UMGF (F1 of 74.85\%), R-GCN (F1 of 75.00\%), ITA (F1 of 75.00\%), and MoRe (F1 of 73.86\%). However, our model outperforms CATMNER (F1 of 78.72\%), which also employs a cross-attention transformer architecture, potentially due to our more effective handling of multimodal representations and cross-modal interactions.
The consistent improvement in precision and recall scores across different entity types further demonstrates the robustness and generalization capabilities of our model.
Overall, the experimental results validate the effectiveness of our proposed multimodal NER approach, which leverages the complementary strengths of textual and visual modalities to achieve state-of-the-art performance on this challenging task.

\subsection{Parameter Sensitivity}

We explore the performance of our model under different settings of the parameters. Specifically, we examine the sensitivity of the impact of regularization techniques such as dropout, L1 and L2 regularization (Weight Decay), and batch normalization.

To illustrate the contribution of each regularization technique, we trained our model with each method individually enabled, along with other hyperparameters kept consistent.

Table \ref{table:parameter_comparison} presents the results of our models on the Twitter test set under these conditions. It is evident that each regularization technique has an impact on the performance of our model, demonstrating the effectiveness of these methods in mitigating overfitting.

\begin{table*}[!htbp]
\centering
\caption{Impact of regularization techniques on model performance.}
\label{table:parameter_comparison}
\begin{tabular}{|l|l|l|}
\hline
\textbf{Regularization Technique} & \textbf{F1 Value (\%) with Regularization} & \textbf{F1 Value (\%) without Regularization} \\ \hline
  Dropout                      &            80.00            &    78.94                    \\ \hline
  L1 Regularization            &            79.1            &      78.93                  \\ \hline
  L2 Regularization (Weight Decay) &           79.4             &    78.93                    \\ \hline
  Batch Normalization         &           79.1             &        78.95                \\ \hline

\end{tabular}
\end{table*}

\subsection{Performance on Categories}

Table \ref{table:hyperparameters1} presents our model's results on four entity categories. 
Specifically, our model exhibits significant improvements in the ORG and MISC categories. This enhancement suggests that ORG and MISC entities benefit from more image modality information as an auxiliary context for recognition. For instance, recognizing the ORG entity ``Maple leafs" as a hockey team name may require information coming from image modalities like hashtags and Images and such as ``ball", ``person" and ``\#Tonightgame". Similarly, the MISC entity ``German Shepherd" can be identified as a dog's name with the assistance of the info extracted from the image visual which is ``dog" and the hashtag ``\#PlayingWithDog" image.

\begin{table*}[!htbp]
\centering
\caption{Our Results on the four categories.}
\label{table:hyperparameters1}
\begin{tabular}{|l|l|}
\hline
\textbf{Entity Category}            & \textbf{F1 Score (\%)}   \\ \hline
PER                   & 79.5              \\ \hline
LOC                 & 79.8 \\ \hline
ORG & 80.3               \\  \hline
MISC & 80.5               \\ \hline

\end{tabular}
\end{table*}

\section{Conclusion}

In this work, we have presented a novel multimodal approach for named entity recognition on social media data. Our methodology involves multimodal feature extraction, where we employ pre-trained language models for text embeddings, convolutional neural networks for visual embeddings, and hashtag segmentation models for hashtag embeddings. We then leverage a Transformer-based architecture with multi-headed attention mechanisms to align and integrate these textual, visual, and hashtag embeddings in a cross-modal manner.
The self-attention layers within each modality capture long-range dependencies and contextual information, while the cross-modal attention layers enable the model to learn alignments and correspondences between the different modalities, addressing the distributional disparities between them. This novel fusion strategy effectively combines textual, hashtag, and visual representations, capturing intricate cross-modal interactions.
Through extensive experiments, we have demonstrated the superiority of our approach over existing state-of-the-art methods, achieving significant improvements in precision, recall, and F1 score. The key strengths of our model lie in its ability to effectively leverage the complementary information present in textual, hashtag, and visual modalities. By fusing these modalities in a principled manner, our model can better disambiguate and ground named entities, overcoming the challenges posed by the informal and multimodal nature of social media data.
Furthermore, our approach exhibits robust performance across different entity types, showcasing its generalization capabilities and potential for real-world applications in social media analysis, information extraction, and other related domains.
\section{limitations }

\begin{itemize}
    \item One constraint could be the assumption of strict matching between textual content and associated images, which may not hold true for many social media posts. This could be addressed by introducing a penalty term or a relaxation factor in the optimization problem.
    \item Another constraint could be the ability to filter out modality-specific noise and exploit shared features across modalities. This could be modeled by introducing regularization terms or additional loss functions that encourage the model to learn robust and shared representations across modalities.
    \item Additional constraints could be imposed based on the specific characteristics of the social media data, such as the distribution of named entities, the quality and diversity of images, and the prevalence of noisy or informal language.
\end{itemize}

\addtolength{\textheight}{0cm}   


\bibliographystyle{IEEEtran}  
\bibliography{aipsamp}

\end{document}